\documentclass [11pt]{article}
\usepackage[dvips]{color}
\usepackage{epsfig}
\def\beq{\begin{equation}}
\def\eeq{\end{equation}}
\def\beqa{\begin{eqnarray}}
\def\eeqa{\end{eqnarray}}
\def\d{{\rm d}}

\def\rc{{\mbox{\tiny \rm c}}}

\def\rL{{\mbox{\tiny \rm L}}}

\def\rmm{{\mbox{\tiny \rm m}}}
\def\rU{{\mbox{\tiny \rm U}}}

\def\rR{{\mbox{\tiny \rm R}}}
\def\ri{{\mbox{\tiny \rm i}}}
\def\rf{{\mbox{\tiny \rm f}}}

\def\rL{{\mbox{\tiny \rm L}}}

\def\rint{{\mbox{\tiny \rm int}}}

\begin{document}
\baselineskip0.55cm plus 1pt minus 1pt
\tolerance=1500

\begin{center}
{\LARGE\bf From mechanics to thermodynamics -- analysis of selected examples}
\vskip0.4cm
{ J. G\"u\'emez$^{a,}$\footnote{guemezj@unican.es},
M. Fiolhais$^{b,}$\footnote{tmanuel@teor.fis.uc.pt}
}
\vskip0.1cm
{\it $^a$ Departamento de F\'{\i}sica Aplicada}\\ {\it Universidad de
Cantabria} \\ {\it E-39005 Santander, Spain} \\
\vskip0.1cm
{\it $^b$ Departamento de F\'\i sica and Centro de
F\'\i sica Computacional}
\\ {\it Universidade de Coimbra}
\\ {\it P-3004-516 Coimbra, Portugal}
\end{center}

\begin{abstract}
We present and discuss a selected set of problems of classical mechanics and thermodynamics.
The discussion is based on the use of the impulse-momentum equation simultaneously  with the centre-of-mass (pseudo-work) equation
or with the first law of
thermodynamics, depending on the nature of the problem. Thermodynamical aspects of
classical mechanics, namely  problems involving non-conservative forces or variation of mechanical energy
are discussed, in different reference frames,  in connection with the use of one or the other energy equation, and with the compliance of the Principle of Relativity.
\end{abstract}

\section{Introduction}
\label{sec:intro}

Relations between mechanics and thermodynamics have been several times pointed out and pedagogically discussed in the literature. In particular we have studied in detail systems whose physics description requires both mechanical and thermodynamical approaches such as the cartesian diver \cite{guemez1} and the
drinking bird \cite{guemez2}. Also many toys, whose physics was revisited in a recent publication  \cite{guemez3}, require a mechanical-thermodynamical analysis in order to fully understand their behaviours.

However, in this work we want to emphasize the use of two ``energy equations", namely the pure mechanical one, relating the variation of the centre-of-mass kinetic energy with the pseudo-work performed by the resultant external force, and the other one, which is simply the equation expressing the First Law of thermodynamics: the variation of the
total energy of a system  is accounted for by all types of works upon the system and by the heat transfers. The level of readership of this article is intended for undergraduate students and our aim is to help students and professors to better address thermodynamical issues in the context of mechanical problems.

The  classical dynamics of a system of  particles of mass $M$ is governed by Newton's Second Law which can be expressed by
\beq
{\vec F}_{\rm ext}\  \d t = M \d {\vec v}_{\rm cm}\, ,
\label{newtonpt}
\eeq
where ${\vec F}_{\rm ext}$ is the resultant of the external forces and ${\vec v}_{\rm cm}$ the center-of-mass velocity. The equation, valid in a given inertial reference frame,  remains invariant upon inertial frame transformations. The previous equation also
allows us to conclude that the infinitesimal variation of the kinetic energy of the centre-of-mass equals the so called
``pseudo-work" \cite{penchina78}. 
Actually, by multiplying both sides of eq.  (\ref{newtonpt}) by ${\vec v}_{\rm cm}$ one concludes that the dot product of the resultant of the external forces by the infinitesimal displacement of the
centre-of-mass is
\beq
{\vec F}_{\rm ext} \cdot \d {\vec r}_{\rm cm} = {1\over 2} M \d v_{\rm cm}^2 \, . \ \label{eq-3}
\eeq
The infinitesimal ``pseudo-work" is the dot product of the resultant force applied to a system and of the corresponding infinitesimal centre-of-mass displacement. It should not be confused with the infinitesimal ``work" of a force which is the dot product of the force and of its own infinitesimal displacement.
The above two equations are equivalent, though they may provide complementary information. The first equation will be referred to as the impulse-momentum equation and states that the infinitesimal impulse of ${\vec F}_{\rm ext}$ equals the infinitesimal variation of the linear momentum of the center-of-mass, $\d {\vec p}_{\rm cm} = M \d {\vec v}_{\rm cm}$. The second equation, which will be referred to as the centre-of-mass equation, states that the infinitesimal pseudo-work equals
the infinitesimal variation of the centre-of-mass kinetic energy, $\d K_{\rm cm}$. Note that, in general, $K_{\rm cm}$ is not the total kinetic energy of the system, $K$, which may also include internal kinetic energy, $K_{\rm int}$, for instance, rotational energy. Of course, from eqs. (\ref{newtonpt}) and (\ref{eq-3}), one readily obtains the corresponding integral equations.

The resultant force acting on the system can be the sum of conservative and non-conservative forces. On the other hand, if there are, for instance, thermal effects, these are not described by the centre-of-mass equation (\ref{eq-3}) \cite{jewett08a,jewett08v}.
Therefore, in order to study the motion of extensive deformable bodies one needs additional equations
to describe rotations \cite{leff93} and processes involving production \cite{mcclelland11} and dissipation  \cite{jewett08d} of mechanical energy. To keep the discussion in this paper at a simple level we do not consider rotations. Here we are particularly interested in
addressing the physical description of the kind of processes
that also need the First Law of thermodynamics \cite{mungan07}.

For any system, its internal energy
variation, $\d U$, receives contributions from the variation of the internal kinetic energy, $\d K_{\rint}$ (including rotational kinetic energy or kinetic energy with respect to the centre-of-mass),
from any internal work, $\d w_{\rint} = -\d \Phi$ ({\em i.e.} work performed by the internal forces) \cite{mallin92},
from the internal energy variations related to temperature variations, $ Mc\d T$ ($c$ is the specific heat),  from the internal energy variations
related to chemical reactions \cite{atkins10}, etc.

In thermodynamics both work and heat contribute to the internal energy variation of a system.
For a general process on a macroscopic deformable body, whose analysis needs to combine mechanics and thermodynamics \cite{jewett08v},
instead of the centre-of mass equation one should consider the  equation
\beq
\d {K}_{\rm cm} + \d { U} = \sum_j { {\vec F}_{{\rm ext}, j}}\cdot \ \d {\vec r}_j  \ + \ \delta { Q}
\label{totale}
\eeq
which is nothing but an expression of the  First Law of thermodynamics \cite{erlichson84,besson01}:
It is more general than the centre-of-mass equation (\ref{eq-3}), but both are valid.
In (\ref{totale}) the infinitesimal heat is denoted by a $\delta $ since it is not an exact differential.
Each term in the sum on the right-hand side of eq.~(\ref{totale})
is work associated with each {\em external} force ${\vec F}_{{\rm ext}, j} = (F_{x, \, j}, F_{y, \, j}, F_{z, \, j})$, and  $\d {\vec r}_j= (\d x_{j}, \d y_{ j}, \d z_{j})$
is the infinitesimal displacement of the  force ${\vec F}_{{\rm ext}, j}$ itself (and not, anymore, the displacement of the centre-of-mass). Therefore, $\delta  {W}_j= {\vec F}_{{\rm ext}, j} \cdot \d {\vec r}_j$ is actual work and not pseudo-work. The above equations are valid in any inertial frame. This means that the amount of information provided by the set of equations is {\em exactly} the same, irrespective of the chosen inertial frame --- there are no privileged inertial frames and the above equations are Galilean invariant.

In many situations there are forces with zero spatial displacement in a certain inertial reference frame (e.g. the force exerted by the ground on the foot of a walking person). In these cases the force produces no work but there is an impulse and a corresponding linear momentum variation: ${\vec F} \d t = \d {\vec p}$.
Let us consider an example from thermodynamics: a gas of small mass (hence, gravitational forces are negligible) is in equilibrium inside a vertical cylinder enclosed by a piston at the top.
A force ${\vec F} = (0, -F, 0)$ is applied to the piston and, hence, by the piston to the gas.
This force is always compensated by an opposite force, $-{\vec F}$, exerted upon the gas by the bottom of the cylinder  (the horizontal forces produced by the cylinder walls have zero resultant and can be ignored). The displacement of this force is always zero.
For a quasi-static process the impulse-momentum equation (\ref{newtonpt}) and the centre-of-mass equation (\ref{eq-3}) applied to this system yields
\beq
\left\{\begin{array}{l}   M \d v_{\rm cm} =  (F-F) \d t \\
{1\over 2} M \d v_{\rm cm}^2 = (F-F) \d y_{\rm cm}\end{array}\right. \, ,
\label{dfcb}
\eeq
leading to $\d v_{\rm cm} = 0$: if the cylinder is initially at rest,  $v_{\rm cm} =0$ during the process, though the
position of the centre-of-mass might change, $\d y_{\rm cm} \not= 0$.

On the other hand, in the energy equation (\ref{totale}), which incorporates
the First Law of thermodynamics,  one has $\d K_{\rm cm}=0$ and, since the bottom force does not do any work, one obtains
\beq
\d U = -F \d y + \delta Q
\eeq
($\d y$ is the displacement of the force exerted by the piston) equivalent to the well known equation $\d U = -P \d V + \delta Q$ (here, $P$ is the pressure and $V$ is the volume).

For a body that behaves like an elementary particle (no rotation, no deformation, etc.) the centre-of-mass equation and the First Law of thermodynamics provide the same information. But when the body does not behave like an elementary particle, such as the gas in the cylinder, the two equations provide complementary information.

Summarizing, besides the impulse-momentum equation, one has at our disposal the centre-of-mass equation and the energy equation \cite{sherwood83}.
Both equations are always valid and, therefore, they should be  compatible. They may provide the same information if the problem under consideration is a purely mechanical one. However, if the situation is within the scope of thermodynamics, the second equation, which embodies the First Law, is more general and provides new information with respect to the centre-of-mass equation.
Of course, one still needs the Second Law of thermodynamics, which states that, observed the First Law, only processes compatible with a non-decrease of the entropy of the universe, $\Delta S_\rU \ge 0$, are allowed.
To illustrate the usage of the impulse-momentum equation together with the centre-of-mass or with the more general energy equation, we discuss, in the next sections, a set of examples ranging from mechanics to thermodynamics. We start with a pure mechanical example, continue with problems involving destruction or production of mechanical energy and conclude with a typical thermodynamical problem.

This paper is organized as follows.
In  section \ref{sec:boai}
we consider the motion of  a block sliding down an incline without friction.
 In section 3 we study a  totally inelastic collision.
In section \ref{sec:boyscat}
 we study an example of mechanical energy production by considering the motion of a person on rollers pushing against a wall.
 In section 5 we treat an inherently inelastic process -- the pulling of a chain.
 Finally, in section \ref{sec:jte}  we address a typical thermodynamical problem -- the Joule-Thomson experiment. With these examples, ranging from a pure mechanical situation to a typical
thermodynamical one, we  show the  appropriateness of the First Law of thermodynamics with respect to the Newton's Second Law in the mechanical description of the system.

\section{Block on an incline}

\begin{figure}[htb]
\begin{center}
\hspace*{-0.5cm}
\includegraphics[width=9cm]{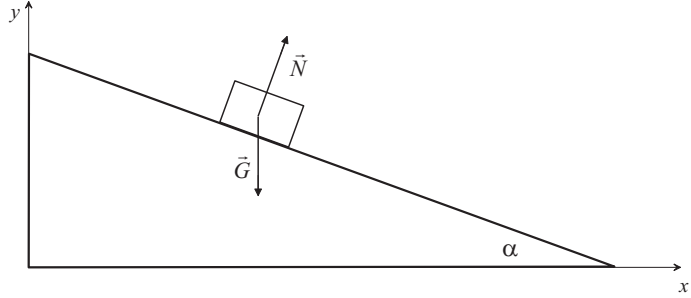}
\end{center}
\vspace*{-0.9cm}
\caption[]{\label{fig:blockinclinenf} \small Rigid  block on an incline.}\vspace*{-0.2cm}
\end{figure}

A rigid  block of mass $M$ slides down an incline without friction, as  Fig.~\ref{fig:blockinclinenf} shows. We describe the process in a reference frame  with horizontal  $x$ and vertical $y$.
\label{sec:boai}
There are two forces, namely the weight, ${\vec G = (0, -Mg, 0)}$ and the normal force,  ${\vec N} = (N \, \sin \, \alpha,  N \, \cos  \, \alpha, 0)$,
where $\alpha$ is the angle between the incline and the horizontal plane. 
The block is a rigid body and there are no rotations, hence the displacement of each force is also the displacement of
the centre-of-mass. The infinitesimal displacements are
\beq
\d {\vec r}_{ G}=  \d {\vec r}_{ N}= \d {\vec r}_{ \rm cm} =
( \d L \, \cos\, \alpha,  -\d L \, \sin \, \alpha , 0 )
\eeq
with $\d L$ the elementary displacement along the incline.
In the absence of rotations or any other change in the internal energy of the body (and no heat transfers)  \cite{bauman92}, the energy equation (\ref{totale}) is exactly the centre-of-mass
equation (\ref{eq-3}).

For the finite process, the block, initially at rest, slides down a distance $L$ during the time interval $t_0$. The integration of the impulse momentum and  of the centre-of-mass equations are straightforward and lead to
\beq
N=Mg\cos \alpha \, ; \hspace*{0.2cm}  t_0   = \left({2   L \over g \, \sin \, \alpha  }\right)^{1/2}\, ; \hspace*{0.2cm} v_{0}   = \left({2 \, L \, g \sin \, \alpha  }\right)^{1/2}\, .
\label{erop}
\eeq
where $v_0$ is the final centre-of-mass velocity. 
These results were obtained in the chosen reference frame, but any other inertial frame could be used, leading to equivalent expressions.
In particular the frame with $x$ along the incline and $y$ perpendicular to it is commonly used.

The simple mechanical process studied in this section is reversible, {\em i.e.}, if the body were launched with opposite velocity from the basis of the incline, it would reach the same original height.
This means that $\Delta S_\rU=0$.  This is not the case if there were kinetic friction and it is also not the case in some of the following examples.

\section{Inelastic collision}

\begin{figure}[bht]
\begin{center}
\hspace*{0.0cm}
\includegraphics[width=13cm] {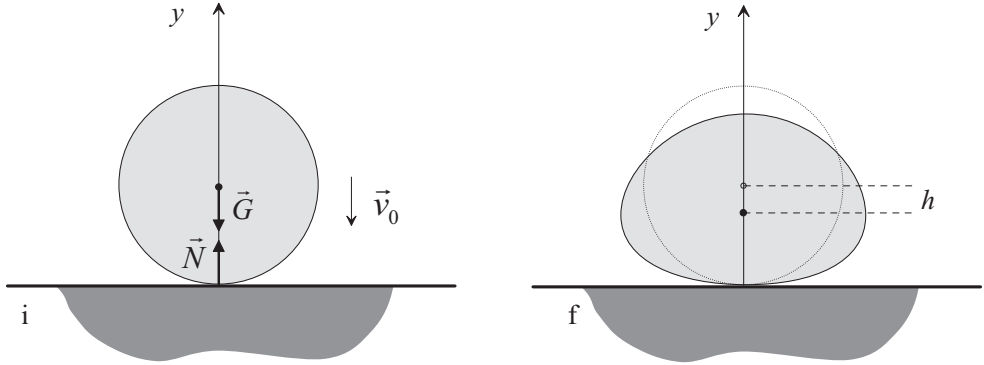}
\end{center}
\vspace*{-0.6cm}
\caption[]{\label{fig:bolachocasuelo} {Inelastic impact of a plasticine ball with the floor (i -- initial contact; f -- ball stops).} }
\end{figure}

The next example  is a perfect inelastic collision of a plasticine ball with the ground~\cite{arons89} and the situation is represented in Fig.~\ref{fig:bolachocasuelo}. We are only interested in the part of the motion after the initial contact with the soil, until the ball
stops completely.

There are two forces: the weight, $\vec G$, and the force, $\vec N$, exerted by the ground, which, for simplicity, is assume to be constant (can be regarded as an average force).
With the $y$ axis directed upwards, the two forces are
${\vec G}= (0, -Mg, 0)$ and $ {\vec N} = (0,N,0)$ .

We denote by $h$  the vertical displacement of the centre-of-mass  and by $t_0$ the time interval since the initial contact of the ball with the floor until
its gets totally stopped.
The displacements associated with the forces and with the centre-of-mass are (note that the displacement of the normal force is zero \cite{leff93}):
\beq
\Delta {\vec r}_G =\Delta {\vec r}_{\rm cm}=  (0, -h, 0) \ \ \ \ \Delta {\vec r}_N = (0,0,0) .
\label{RGN}
\eeq

After integration, the centre-of-mass equation (\ref{eq-3}) leads to
$\Delta {K}_{\rm cm}  = \left({\vec G} + {{\vec N}}\right)\cdot  \Delta {\vec r}_{\rm cm}$ and, denoting by $v_{\rm cm, i}=-v_{0}$ the velocity of the centre-of-mass when the ball reaches the floor,
one obtains ($v_{\rm cm, f}=0$), for the impulse-momentum and center-of-mass equation,
\beq
\left\{
\begin{array}{rl}
M v_{0} &=  (N - Mg)  t_0 \, , \vspace*{0.3cm}\\
{1\over 2} M v_{0}^2 &= (N - Mg) h \, .
\end{array}
\right.
\label{850o}
\eeq
Note that the second equation, namely
${1\over 2} M v_{0}^2 + M g h =   N h$, is not a  `work-energy' equation since ${ N} h$ is pseudo-work.

Let's now exploit the energy equation (\ref{totale}), which can also be readily integrated (forces are constant), assuming the form
\beq
\Delta { K}_{\rm cm} + \Delta { U} = {\vec G}\cdot \Delta {\vec r}_G + {\vec N}\cdot \Delta {\vec r}_N + { Q}\, .
\label{87uy}
\eeq
In the absence of other effects (rotations, work of internal forces) the variation of the
internal energy of the ball is assumed to be
\beq
\Delta { U}  = M c (T_{\rm f} - T_{\rm i}) \, ,
\eeq
where $T_{\rm i,f}$ are the initial and final temperatures of the ball and $c$ its specific heat. This and $Q$  in (\ref{87uy}) accounts for thermal effects in this type of inelastic collisions.
The impulse-momentum equation (\ref{newtonpt}) and eq.~(\ref{87uy}) allow us to write
\beq
\left\{
\begin{array}{rl}
M v_{0} &=  ( N - Mg)  t_0\, , \vspace*{0.3cm}\\
-{1\over 2}   M v_{0}^2 + M c (T_{\rm f} - T_{\rm i})  &= M g h + Q
\, .
\end{array}
\right.
\label{eq:pbcnsl}
\eeq
To experimentally confirm these equations, now one would need a thermometer, in addition to a watch  and a ruler.

The two sets of equations (\ref{850o}) and (\ref{eq:pbcnsl}) are both valid and must be compatible.
From the second equations of each set one concludes that $Mc(T_{\rm f} - T_{\rm i}) = Q+ N\, h$ \cite{koser11}.
Assuming that the temperature change of the ball
is negligible, $T_{\rm f} \simeq T_{\rm i}\simeq T$, one concludes that there is an energy transfer, $Q=-Nh$, from the system.
This energy is thermal energy (heat) exchanged with the thermal reservoir at temperature $T$ and, therefore, it is a positive quantity from  the reservoir point of view.
The variation of the entropy of the universe is solely the variation of the entropy of the thermal reservoir and it is given by
\beq
\Delta S_\rU ={Nh \over T} > 0\, .
\eeq
For a deformed ball initially at rest on the ground, one could think about an injection of the amount of energy $N\, h$ from the heat reservoir, e.g. through the action of the force $N$.
This process, opposite to the one that we have studied, would be allowed by the First Law of thermodynamics. However, it would lead to a decrease of the entropy of the universe,
which is prohibited by the Second Law of thermodynamics \cite{smith83}.
Therefore this inverse process cannot happen spontaneously \cite{leff12b}.
This is not the case for the pure mechanical example considered in the previous section, which does not increase the entropy.

Let us now consider the description of the same inelastic process in the reference frame S$'$, moving vertically, along the common $y'y$ axes, with velocity $V$  with respect to S.

The forces are identical in both reference frames but  the displacements in S$'$, $\Delta {\vec r\, }'$, differ from (\ref{RGN}) in the $y$ component, where there is an extra term $-V t_0$:
\beq
\Delta {\vec r\, }'_G =\Delta {\vec r\, }'_{\rm cm}=  (0,  -h-V t_0, 0)\, , \ \ \ \
\Delta {\vec r\, }'_N = (0, -V t_0, 0)
\eeq
The impulse-momentum and the center-of-mass equations in the new frame are
\beq
\left\{
\begin{array}{rl}
M v_{0} &=  (N - Mg)  t_0\, , \vspace*{0.3cm}\\
Mv_{0} V+{1\over 2} M v_{0}^2 &= (N - Mg) (h+Vt_0) \, .
\end{array}
\right.
\label{850pl}
\eeq
which reduces exactly to (\ref{850o}). Formally, in any situation, the equations in reference S$'$ are always linear combinations of the equations in reference S, and vice-versa, according to the Principle of Relativity, and we have explicitly shown this for a particular case. Therefore, by changing to a new inertial reference frame,
one does not gain or loose information (though one reference frame might be better suited than another one regarding technical aspects)

Next we explore the energy equation also in S$'$:
\beq
\Delta { K}'_{\rm cm}  + \Delta {U}' = { W}'_{G} + {W}'_{N} + {Q}'\, ,\nonumber
\eeq
The internal energy variation and the heat are the same in both frames: $\Delta { U}'=\Delta { U}$
and ${ Q}'={ Q}$. The impulse-momentum and the energy equation are readily obtained and
altogether lead to
\beq
\left\{
\begin{array}{rcl}
- M V + M (v_{0} + V)  &=& (N - M g) t_0\,\vspace*{0.3cm} \\
-{1\over 2}   M v_{0}^2 - M V v_{0} + M c (T_{\rm f} - T_{\rm i})  &=& M g h + M g V t_0-  N V t_0  +Q \, .
\end{array}
\right.
\label{eq:pbfltsb}
\eeq
Again, these equations reduce to (\ref{eq:pbcnsl}) as expected. The discussion on thermal effects, heat transfer and increase in the entropy of the universe,
presented above, still applies here --- the temperature and the entropy variation,  $\Delta S_\rU$, are
Galilean invariants.
It is worth noticing that eq.~(\ref{totale}) can still be written as $\d K_{\rm cm} - \sum_j { {\vec F}_{{\rm ext}, j}}\cdot \ \d {\vec {r}}_j= \d U - \delta Q$ and both sides of the equation are, {\it per se}, Galilean invariant too.

\section{Mechanical energy production}
\label{sec:boyscat}

\begin{figure}[htb]
\begin{center}
\hspace*{-1.3cm}
\includegraphics[width=9.0cm]{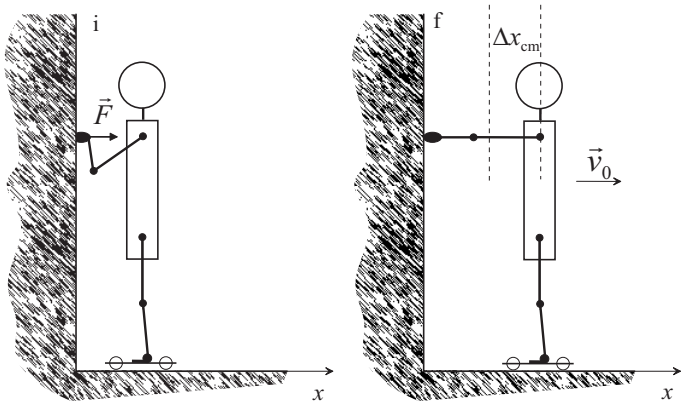}
\end{center}
\vspace*{-0.7cm}
\caption[]{\label{fig:personscaters} \small Person on rollers pushing against a wall. }\vspace*{-0.1cm}
\end{figure}

A boy on rollers, initially at rest, pushes against a wall (with infinite mass) and slides away over the floor \cite{mcclelland11}.
The mass of the system is $M$ and no dissipative forces between the rollers and the floor are considered. An external horizontal force, $\vec F$, is exerted on him by the wall and, as a consequence, his translational kinetic energy increases.
The vertical forces --- the weight and the normal force from the ground --- cancel each other and they play no role in the discussion.
The force $\vec F$ does not do any work because its displacement is zero \cite{arons89} (Fig.~\ref{fig:personscaters}).

This is again a one dimensional problem and we describe it in reference frame S with the $x$ axis pointing to the right.  The force along the $x$ direction is ${\vec F} = (F, 0, 0)$, represented in the figure.
We assume $F$ to be constant ($F$ can be considered an average force).
The displacements of the centre-of-mass and of the horizontal force are
\beq
\Delta {\vec r}_{\rm cm}  = (\Delta x_{\rm cm}, 0 , 0)\, , \ \ \
\Delta {\vec r}_F = (0,0,0) \,
\eeq
where $\Delta x_{\rm cm}$ is the spatial displacement of the centre-of-mass occuring in the time interval $t_0$,  the duration of the action of the horizontal force upon the person.
Denoting by $v_0$ the velocity of the boy's centre-of-mass when the
contact with the wall ceases, the impulse-momentum and the centre-of-mass equation lead to
\beq
\left\{
\begin{array}{rl}
M v_{0} &=  F  t_0 \, , \vspace*{0.3cm}\\
\ \ {1\over 2} M v_{0}^2 &= F \Delta x_{\rm cm} \, .
\end{array}
\right.
\label{8670}
\eeq
We stress that the second equation is not an expression of an `energy-work' theorem since the external normal force exerted by the wall does not produce any work~\cite{dufresne11}. If the force does not do any work, where is the source of energy responsible for the increase of the centre-of-mass kinetic energy?

These energetic issues are better discussed using the energy equation.
During the process of pushing against the wall, the work of $\vec F$ vanishes but not its impulse. There are biochemical reactions in the person's muscles.
When a chemical reaction $\xi$ is produced inside the body, these may cause variations in its internal energy, $\Delta U_\xi$, volume, $\Delta V_\xi$, entropy, $\Delta S_\xi$, etc.
If a chemical reaction takes place at constant external pressure, $P$, in diathermal contact with a heat reservoir at temperature $T$, part of the internal energy is used for the  expansion against the external pressure and part must be exchanged with the heat reservoir in order to ensure that the entropy of the universe does not decrease. The  work, ${W}_P$, and the heat, ${Q}_\xi$,  are:
\beq
{W}_P =    - P \Delta V_\xi  \, ; \ \ \   {Q}_\xi =    T \Delta S_\xi \, ,
\eeq
When $\Delta V_\xi < 0$, the external pressure performs work on the system and when $\Delta S_\xi > 0$ the heat reservoir
increases the internal energy of the system. We assume that no other  energy is exchanged between the person and the environment (no extra heat is exchanged between the person and the environment).

The energy equation for this process --- eq. (\ref{totale}) --- is:
\beq
\Delta { K}_{\rm cm} + \Delta\,  {U}_\xi  = {W}_{F} + {W}_{P} + { Q}\, ,
\eeq
and, together with the impulse-momentum equation,
\beq
\left\{
\begin{array}{rl}
 M v_{0} &= F  t_0 \, , \vspace*{0.3cm}\\
{1\over 2} M v_{0}^2 + \Delta U_\xi &= -P \Delta V_\xi + T\Delta S_\xi\, .
\end{array}
\right.
\label{eq:fltboys}
\eeq
The second equation provides new information with respect to the second equation in (\ref{8670}).
In particular, the second equation in eq.~(\ref{eq:fltboys}) can also be expressed as
\beq
{1\over 2} M v_{0}^2  = - \Delta G_\xi\, ,
\eeq
where:
\beq
\Delta G_\xi = \Delta U_\xi +P \Delta V_\xi - T\Delta S_\xi\, ,
\eeq
is the Gibbs free energy variation, which is symmetric of the pseudo-work associated with $\vec F$. This equation
shows that, in effect, the person increases its centre-of-mass kinetic energy thanks to the internal biochemical reactions \cite{atkins10}.
For given temperature and pressure, the variation of the Gibbs free energy is the maximum useful work that can be obtained: $W_{\rm max} = -  \Delta G_\xi$.

In the reference frame S$'$, in standard configuration with respect to S,
the centre-of-mass equation is
\beq
 \Delta {K}'_{\rm cm}   = {\vec  F}\,\cdot \Delta{\vec r\, }'_{\rm cm}\, ,
\eeq
which, again with the impulse-momentum equation results in  (in S$'$ the displacement of the force exerted on the person's hand is $-Vt_0$)
\beq
\left\{
\begin{array}{rcl}
- M V - M (v_{0}-V)  &=&  F  t_0 \, , \vspace*{0.3cm}\\
{1\over 2} M V^2 - {1\over 2} M (v_{0}-V)^2 &=& F \left(\Delta x_{\rm cm}- V t_0\right)\, ,
\end{array}
\right.
\eeq
equivalent to (\ref{8670}). The same happens with the energy equation in reference frame S$'$: from
$\Delta { K}'_{\rm cm} + \Delta\,  { U'}_\xi  = { W'}_{F} + {W'}_{P} + {Q'}\, ,$
one obtains (\ref{eq:fltboys}).

The process studied in this section implies no entropy increase of the universe and, therefore, it is reversible. In fact, the kinetic energy of the system can be completely used, at least in principle,  to increase by $-\Delta G_\xi$ the free energy of any chemical reaction.
Real bodies, which are articulated or made of elastic materials, may acquire accelerations (changes of centre-of-mass momentum and kinetic energy) as a consequence of the action of internal forces when they are in contact with an infinite mass body (a wall, the floor, etc.); of course, internal forces, by themselves, cannot accelerate the center-of-mass of an isolated body \cite{bernard84}.
The final kinetic energy  in the example studied in this section, or in the case of a person that jumps,
is not due to the work of the contact forces since they have zero displacement \cite{mcclelland85}.
Any explicit or implicit assumption that forces  with zero displacement applied to a moving body (forces on the foot of a person that walks, on the tires of a car that accelerates, on the foot of a person that jumps, etc.) do work is incorrect.
In fact, the total kinetic energy of a composite, articulated body in contact with an infinite mass body can change, even when no work is done by the external forces applied to it.
In the description of processes with production of mechanical energy,  the role of the chemical reactions
(food  consumption in people that walk or jump, fuel combustion in cars, etc.)
plays a central role.

In this section we analised a case of mechanical energy production. In examples with mechanical energy destruction (e.g. the plasticine ball
that stops and deforms after its impact with the floor, or the inherently elastic process of pulling a chain to be treated in the next section)
we come out to the conclusion that the entropy of the universe increases.

\section{Pulling a chain}

\begin{figure}[htb]
\begin{center}
\hspace*{0.0cm}
\includegraphics[width=12cm] {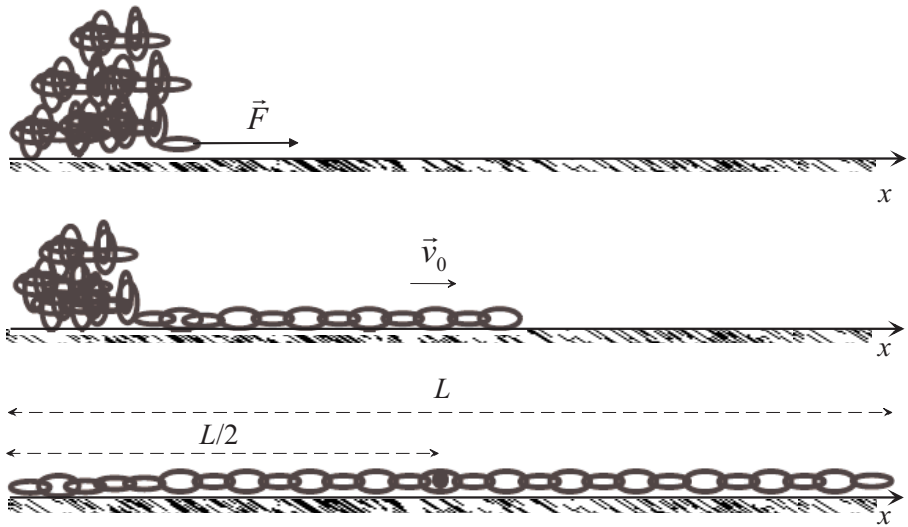}
\end{center}
\vspace*{-0.6cm}
\caption[]{\label{fig:cadenaquesedesenrolla} A chain unwraps under the action of a force  $\vec F$ moving with constant velocity $v$. }
\end{figure}

A chain of length $L$ and mass density $\lambda$ lies in a heap on the floor \cite{sousa02}.
Its end is grabbed and pulled horizontally with a force $\vec F$, as Fig.~\ref{fig:cadenaquesedesenrolla} shows.
As a result,  every piece of the moving unwrapped chain acquires a {\em constant} velocity ${\vec v}_0$.

We denote by $x$ the position of the end of the chain, so that $L-x$ is the length of the chain that remains on the heap.
The centre-of-mass of the chain is
located at position (we assume that the heap is narrow)
\beq
x_{\rm cm} = {\lambda x (x/2)\over \lambda L} = {x^2\over 2L}\, .
\eeq
The centre-of-mass velocity is the time derivative and it is given by
\beq
v_{\rm cm} = {x v_0\over L}\, .
\eeq
The centre-of-mass momentum (product of the total mass by the centre-of-mass velocity) is
\beq
p_{\rm cm} = \lambda L \, v_{\rm cm} = \lambda    x\, v_0=p\, .
\eeq
meaning that the linear momentum of the moving part of the chain is always equal to the linear momentum of the
centre-of-mass.
The centre-of-mass kinetic energy is (half of the total mass times $v_{\rm cm}^2$)
\beq
K_{\rc \rmm }= {1\over 2} \lambda L v_{\rm cm}^2 = {1\over 2} \lambda {x^2 v^2_0  \over L}\, ,
\eeq
which is, in general, different from the total kinetic energy given by (half of the mass of the already moving part times $v^2_0$) :
\beq
K = {1\over 2} \lambda x v^2_0\, .
\eeq

When the end of the chain moves by $\d x$, the above mentioned magnitudes change accordingly, and their infinitesimal changes are given by
\beqa
\d x_{\rm cm} &=&   {x \over  L}\,\d x\, ,\\
\d v_{\rm cm} &=& { 1  \over L} v _0\,\d x\, ,\\
\d p_{\rm cm} &=&  \lambda   {  v}_0 \,\d x = \d p \, , \label{eqpa}\\
\d K_{\rc \rmm }&=&     \lambda {x    \over L} v^2_0 \, \d x\, , \label{eqpx}\\
\d K &=& {1\over 2} \lambda   v^2_0 \, \d x\, .\label{eqpb}
\eeqa
This is a one dimensional motion and the force is given by
 ${\vec F}=(F, 0, 0)$. Its infinitesimal displacement and the infinitesimal centre-of-mass displacement, occurring in time $\d t$ are
 \beq
  \d {\vec r}_F = (\d x, 0, 0) \, \ \ \ \
\d {\vec r}_{\rm cm} = \left( {x \over  L}\d x, 0, 0\right) \, .
\eeq
The infinitesimal variation of the linear momentum and of the centre-of-mass kinetic energy  [see eqs. (\ref{eqpa}) and~(\ref{eqpx})]
immediately lead to  the following set of equations
\beq
\left\{
\begin{array}{rcl}
\lambda  v_0 \d x &=& F \d t \vspace*{0.3cm}\\
\lambda {x \over L} v^2_0 \d x &=& F {x \over L} \, \d x\, .
\end{array}
\right.
\label{yu70}
\eeq
Finally, one concludes that the pulling force must be constant and given by \cite{morin08}
\beq
F = \lambda v^2_0 \, .
\label{forcav2}
\eeq

The  energy equation is
$\d { K}_{\rm cm} + \d { U} = {\vec F}\cdot \d {\vec  r}_F  + \delta { Q}$.
If one assumes that the chain does not change its temperature during the process,
the internal energy is then just the internal kinetic energy, {\em i.e.} the kinetic energy relative to the centre-of-mass) and therefore
$\d U= \d { K}- \d { K}_{\rm cm}$ (K\"oning's theorem \cite{knudsen00}).
Hence
\beq
\d {K}= {\vec F} \cdot \d {\vec  r}_F  + \delta { Q}
\eeq
where $\d {K}$ is the kinetic energy variation.
This equation, together with the impulse-momentum equation  [see (\ref{eqpa}) and (\ref{eqpb}) ]
allows us to arrive at
\beq
\left\{
\begin{array}{rl}
 \lambda  v_0 \, \d x &= F \, \d t \vspace*{0.3cm}\\
{1\over 2} \lambda   v^2_0 \, \d x &= F\,  \d x  + \delta Q \, .
\end{array}
\right.
\label{yu71}
\eeq
From the first equation we obtain the information already included in (\ref{yu70}): with $\d x =v_0 \d t$, one gets (\ref{forcav2}).
From the second equation in (\ref{yu71}) one obtains
\beq
{1\over 2} \lambda   v^2_0 \, \d x = \lambda v^2_0 \, \d x  + \delta Q\, .
\eeq
Therefore, thermal effects are accounted for through the equation:
\beq
Q  = \int \delta Q = - {1\over 2} \lambda v^2_0 \int_0^x \, \d x = - {1\over 2} \lambda v^2_0 x = -{1\over 2} F x.
\eeq
This means that half of the work, $F x$, produced by the pulling force is dissipated as thermal energy, with the corresponding increase of the entropy of the universe.

The displacement of chains --- articulated bodies formed by rigid links and with no internal sources of  energy ---, is an inherently irreversible process.
There are links of the chain that vary their speeds abruptly, and these  processes destroy mechanical energy \cite{sherwood83}.
Each link in the chain that is, abruptly, placed into motion can be regarded as an inelastic collision, producing a loss of mechanical energy and making the process irreversible \cite{morin08}.

The description of the process in reference S$'$ in standard configuration is also very instructive for the  illustration of  the Principle of Relativity at work.

\section{The Joule-Thomson experiment}
\label{sec:jte}

\begin{figure}[htb]
\begin{center}
\hspace*{-0.5cm}
\includegraphics[width=9cm]  {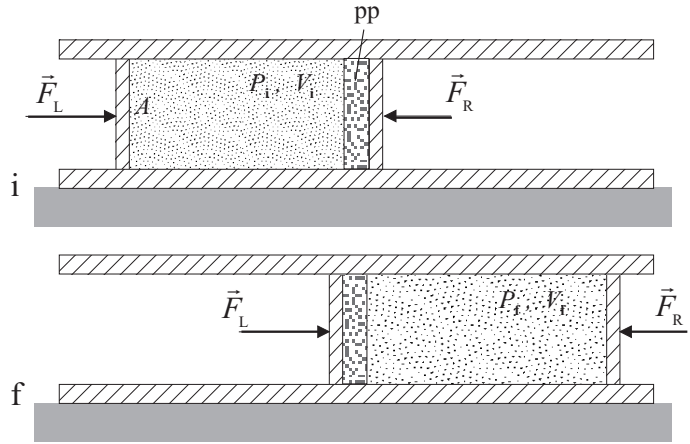}
\end{center}
\vspace*{-0.5cm}
\caption[]{\label{fig:joth-tr-fig01}  \small The Joule-Thomson or porous-plug (pp) experiment.}\vspace*{-0.2cm}
\end{figure}

William Thomson (who became Lord Kelvin) proposed a method to study  the dependence of the internal energy of a gas with its volume.
Together with Joule, they carried out a series of experiments using a device sketched in  Fig.~\ref{fig:joth-tr-fig01} \cite{zemansky97}.
A cylindrical tube, with cross-section of area $A$, is divided into two parts by a porous-plug.
In the Joule-Kelvin (or porous plug) process a gas (usually a real gas), of mass $M$, occupying the initial volume $V_\ri$, on the left hand side,
and pressure $P_\ri$, is forced to pass through a porous plug or a valve against a final constant pressure $P_\rf<P_\ri$, reaching a final volume $V_\rf$, on the right hand side.
During the process the pressures on the left and on the right (see figure) are  kept constant but the gas temperature may change from $T_\ri$ to $T_\rf$.

There are three forces acting on the gas: the left side force,  ${\vec F}_\rL=( A P_\ri, 0, 0)$,
the right side force,  ${\vec F}_\rR=(- A P_\rf, 0, 0)$, and the zero displacement force, ${\vec N}=(N,0,0)$, exerted on the gas by the  porous plug \cite{penn67}.
The displacements are
\beq
\Delta {\vec r}_\rL= ( V_\ri / A  , 0, 0) \, ,  \ \ \Delta {\vec r}_\rR= (  V_\rf / A  , 0, 0)  \, , \ \
\Delta {\vec r}_N= (0,0,0)\, , \ \
\Delta {\vec r}_{\rm cm}= (  \Delta x_{\rm cm}, 0,0)\, .
\eeq
We denote by $\Delta x_{\rm cm}$  the horizontal displacement of the centre-of-mass, $\Delta x_{\rm cm}=(V_\ri+V_\rf)/(2A)$ during the time interval $t_0$, the duration of the process. For the sake of generality let  us consider initial and final centre-of-mass velocities denoted by $v_{\rm i,f}$
(in a more restricted analysis they could be simply set to zero). The impulse momentum equation, $({\vec  F}_\rL+{\vec  F}_\rR+{\vec N}) t_0 = \Delta {\vec p}_{\rm cm}$  and the centre-of-mass equation, $\Delta { K}_{\rm cm}= ({\vec  F}_\rL+{\vec  F}_\rR+{\vec N}) \cdot \Delta {\vec r}_{\rm cm}$ yield
\beq
\left\{
\begin{array}{rl}
M \left(v_{\rm f} - v_{\rm i}\right) &=   \left(P_\ri A - P_\rf A + N\right)  t_0 \vspace*{0.3cm}\\
{1\over 2} M  \left(v_{\rm  f}^2 - v_{\rm  i}^2\right)  &= (P_\ri A - P_\rf A + N ) \Delta x_{\rm cm}\, .
\end{array}
\right.
\label{yu70yh}
\eeq
If the initial and final centre-of-mass velocities are zero, both equations lead to the same equation for $N$, namely
\beq
 N= (P_\rf  - P_\ri) A
 \label{nvbg}
\eeq
{\em i.e.} the force exerted upon the gas by the  porous-plug is directed to the left because $P_\rf  < P_\ri$.

Now we consider the energy equation,
$
\Delta { K}_{\rm cm} + \Delta { U }=  {\vec  F}_\rL \cdot \Delta {\vec r}_{\rL} + {\vec  F}_\rR \cdot \Delta {\vec r}_{\rR}+ {\vec N} \cdot  \Delta {\vec r}_N + { Q}\, ,
$
and assume zero centre-of-mass velocity \cite{tykodi67},
hence eq. (\ref{nvbg}) holds. In such case,
\beq
\left\{
\begin{array}{rl}
0 &= \left(P_\ri A - P_\rf A + N\right) t_0 \vspace*{0.3cm}\\
\Delta U  &= P_\ri V_\ri - P_\rf V_\rf + Q\, .
\end{array}
\right.
\label{yu70yo}
\eeq

The Joule-Thomson experiment is carried on under adiabatic conditions, {\em i.e.} the device (cylinder and pistons) is thermally isolated so  there is  no heat involved in the process.
In this case the second eq. (\ref{yu70yo}) reduces to

\beq
U_\rf- U_\ri  = P_\ri V_\ri - P_\rf V_\rf
\eeq
and, since the enthalpy is related to the internal energy through $H=U+PV$, one concludes that the process is isoenthalpic:  $H_\rf=H_\ri$.

We can finally show that the entropy of the universe increases in this irreversible process. From $\d H= V \d P + T \d S$, for an isoenthalpic process
\beq
\d S_H = -{V\over T} \ \d P_H > 0
\eeq
since  $\d P_H$ is always negative (the pressure decreases in the process).
The Joule-Thomson experiment is a typical thermodynamical process in which all external forces have zero resultant (and information resulting from the centre-of-mass equation).

If the process is studied in reference frame S$'$, moving with velocity $V$ along the horizontal direction, the porous-plug moves with velocity $-V$ and the result set of equations is
\beq
\left\{
\begin{array}{rl}
0 &= \left(P_\ri A - P_\rf A + N\right) t_0\vspace*{0.3cm}\\
\Delta U  &= P_\ri V_\ri - P_\rf V_\rf - V\left(F_{\rm L} - F_{\rm R} +N\right) t_0 +   Q\, ,
\end{array}
\right.
\label{yu70yosp}
\eeq
totally compatible with the equations in reference S.

\section{Conclusions}

The centre-of-mass equation, resulting from the integration of the Newton's Second Law,
and the First Law of thermodynamics, establishing the energy balance in a process, correspond to two absolutely different physical hypothesis.
Both equations are valid but their applicability deserves special attention. We addressed this issue, in a pedagogical way, by firstly presenting the equations and
then applying them to selected examples. Situations for which there is mechanical energy production are particularly interesting. These include, among others,  problems
related with human activities (a person jumping, climbing, cycling, etc.)
or with living beings (a horse pushing  a chariot, etc.) and with thermal engines (a car that accelerates due to an internal combustion, etc.),
which are sometimes incompletely or even incorrectly addressed. This was actually a motivation for the present work where we have discussed interesting physical
situations ranging from mechanics to thermodynamics.

 Together with the impulse-momentum vector equation we used a scalar energy equation, which is either the centre-of-mass equation or the First Law of thermodynamics,
emphasizing the additional information that can be obtained from the second one with respect to the other.
We also paid some attention to the frame transformation: the set of equations appropriate to discuss a problem provide the same information irrespective of the chosen inertial reference frame (Principle of Relativity).
This was illustrated using various examples. Of course, the choice of the reference frame is usually determined by the simplicity of the resulting formalism.

 Intrinsic dissipative processes, such as processes with chains, should be related with thermodynamics when they are solved. When a chain is unwrapped or raised, a  significant  part of the work provided by the external force that displaces the chain is dissipated as heat, an effect that deserves formalization into an equation, and which is not included when attention is only paid to the mechanical part of the problem.
We also considered a couple of examples involving forces acting without displacing their application points. In a certain reference frame, such a zero displacement force produces no work but it does have an impulse.
In another inertial frame there might be both impulse and work.
Whenever pertinent we discussed other thermodynamical aspects, namely the irreversibility of some processes in connection with the Second Law.
On the contrary, we have discussed the forces that act in thermodynamical processes, something that is really uncommon in thermodynamical textbooks. For instance in the Joule-Kelvin process the porous plug exerts a force on the gas and, even though it does not do any work, it plays a crucial role in the description of the process.

Through the discussion of the five examples presented in this article we aimed at clarifying the additional information provided by the more general First Law of thermodynamics with respect to
Newton's Second Law.

\end{document}